\documentclass[a4paper,twoside]{article}

\usepackage{epsfig}
\usepackage{subcaption}
\usepackage{calc}
\usepackage{amssymb}
\usepackage{amstext}
\usepackage{amsmath}
\usepackage{amsthm}
\usepackage{multicol}
\usepackage{xcolor}
\usepackage{pslatex}
\usepackage{apalike}
\usepackage{algorithm2e}
\usepackage{titling}
\usepackage[bottom]{footmisc}
\usepackage{rotating}
\usepackage{hyperref}
\usepackage{SCITEPRESS}     

\begin{document}


\title{Social Norm Reasoning in Multimodal Language Models: An Evaluation}

\author{\authorname{Oishik Chowdhury\sup{1}, Anushka Debnath\sup{1} and Bastin Tony Roy Savarimuthu\sup{1}}
\affiliation{\sup{1}School of Computing, University of Otago, New Zealand}
\email{oishik.chowdhury@postgrad.otago.ac.nz, anushka.debnath@postgrad.otago.ac.nz, tony.savarimuthu@otago.ac.nz}
}


\abstract{
In Multi-Agent Systems (MAS), agents 
are designed with social capabilities, allowing them to understand and
reason about social concepts such as norms when interacting with others (e.g., inter-robot interactions). In Normative MAS (NorMAS), researchers study how norms develop, and how violations are detected and sanctioned. However, existing research in NorMAS use symbolic approaches (e.g., formal logic) for norm representation and reasoning whose application is limited to simplified environments. In contrast, Multimodal
Large Language Models (MLLMs) present promising possibilities to develop software used by robots to identify and reason about norms in a wide variety of complex social situations embodied in text and images. However, prior work on norm reasoning have been limited to text-based scenarios. This paper investigates the norm reasoning competence of five MLLMs by evaluating their ability to answer norm-related questions based on thirty text-based and thirty image-based stories, and comparing their responses against humans. Our results show that MLLMs demonstrate superior performance in norm reasoning in text than in images. GPT-4o performs the best in both modalities offering the most promise for integration with MAS, followed by the free model Qwen-2.5VL. Additionally, all models find reasoning about complex norms challenging.}


\onecolumn \maketitle \normalsize \setcounter{footnote}{0} \vfill

\section{INTRODUCTION\footnote{\color{red}{This paper appears in the proceedings of ICAART 2026.}}}
\label{sec:introduction}

Norms define behavioral expectations that are essential for maintaining social order within the society \cite{durkheim2016division,conte2014minding}. They direct acceptable actions which enable coordination and cooperation between agents, leading to stability in multiagent environments \cite{boella2006introduction}.
A norm-capable software agent should be able to identify relevant norms, detect compliance or violations, and make informed decisions about whether to follow or violate these norms, considering the consequences. Such decisions are often shaped by the agent’s goals, social context, and perceived trust or reputation \cite{savarimuthu2011norm,castelfranchi2000trust}.

The field of Normative Multi-Agent Systems (NorMAS) studies these capabilities, including norm emergence and enforcement via sanctioning \cite{hollander2011current,morris2019norm}. Many works often use symbolic reasoning methods based on deontic logic \cite{boella2006introduction}. Although powerful, symbolic reasoning approaches like deontic logic often face challenges with scalability and adaptability in dynamic, real-world settings. Moreover, norms in these approaches have to be manually encoded into logic based on natural language specifications. Large Language Models (LLMs) offer a promising alternative without the need for encoding norms into formal specification and also enable context-sensitive reasoning about norms across varied social situations \cite{he2024norm}. Emerging research is exploring how LLMs can enhance MAS architectures by improving norm recognition and adaptability \cite{warnakulasariya2025,he2024norm}.

Recent advancements in LLMs have demonstrated rich contextual understanding and inference abilities in social and ethical reasoning tasks \cite{achiam2023gpt}. Their capacity to process complex inputs, including natural language and visual cues, makes them ideal candidates for assessing norm reasoning (adherence or violations). In particular, multimodal LLMs (MLLMs), that  integrate textual and visual modalities, present a powerful approach for analyzing socially grounded behaviors and interactions in rich, real-world-inspired settings \cite{tsimpoukelli2021multimodal}. However, no prior work has investigated norm reasoning abilities of MLLMs.

Our work bridges the gap by introducing a comprehensive evaluation framework designed to systematically assess normative reasoning capability of MLLMs based on normative stories presented in the form of text and images. 30 text-based stories representing five norms, each with six variants (5 * 6 = 30) were presented to five different MLLMs (GPT-4o, Gemini 2.0 Flash, Qwen-2.5VL (72B), Intern-VL3 (14B), and Meta LLaMa-4 Maverick) and their answers to eight questions for each story was analysed based on comparing against ground truth from human evaluation. Similarly, 30 image-based stories were analysed by the same five MLLMs, and the result was compared against  human evaluation.

Ultimately, our broader vision is to enable the development of socially intelligent software agents such as social robots or mobile apps that can be used to identify scenarios where certain norms apply and whether someone has violated the norm. Rather than relying on explicit rule encoding for every possible social situation, we envision robots that can learn, and reason about social norms based on interaction histories of agents, and use LLMs or MLLMs as the engine for norm learning and reasoning. This motivation aligns closely with the goals of NorMAS, where agents must not only act autonomously but also behave in socially compliant, context-aware ways. Our work contributes towards this vision by evaluating whether MLLMs can serve as foundational components in enabling such agents to interpret and adhere to social norms from diverse inputs, including natural language-based inputs (i.e., text inputs) and visual scenes (i.e., image inputs).

\section{\uppercase{Prior Work}}

Research on norms in MAS and LLM environments has recently converged around two complementary themes: extracting and reasoning with explicit norms, and the spontaneous emergence of norms in agent societies. Our work falls under the first theme. Below, we organize prior work into two primary threads: 1) MAS research that focuses specifically on norms (i.e., NorMAS) and 2) NLP research, each contributing uniquely to the understanding of norms.

\textbf{NorMAS research} - This branch explores how norms can be explicitly represented, enforced, and reasoned about within artificial agent societies. Several researchers have discussed foundational MAS norm theories, the use of norm life cycle models \cite{savarimuthu2011norm,frantz2018,morris2019norm}, and the role of meta-norms \cite{mahmoud2015}, thus establishing a theoretical base for norm reasoning within agent systems. Traditional research in MAS has emphasized embedding norms into artificial societies using symbolic methods, but recent developments have shifted toward more adaptive, generative paradigms. Savarimuthu et al. have advocated the use of LLMs as normative reasoning engines within MAS, facilitating norm discovery, reasoning, and compliance while addressing the rigidity of earlier symbolic methods \cite{savarimuthu2024harnessing}. Ren et al. propose the CRSEC framework  (Creation, Representation, Spreading, Evaluation, and Compliance), for inducing norms in sandbox environments, and showing how norms help mitigate inter-agent conflict \cite{ren2024emergence}. Li et al. \cite{li2024agent} explore how agents align with and adapt to evolving social norms, offering methods to quantify this alignment. Extending this trajectory, Warnakulasuriya et al. \cite{warnakulasariya2025} investigate norm emergence and cooperation in LLM-agent societies by analyzing the role of different punishment strategies in promoting cooperative behavior, marking a novel contribution to the NorMAS community through the integration of LLMs and norm-enforcement dynamics.

\textbf{NLP research} - Parallelly, the NLP community has investigated how LLMs interpret, extract, and adapt to norms primarily within textual environments. Haque et al. explore legal norm extraction using ChatGPT, highlighting the mapping of commitments, obligations, and prohibitions, while also exposing issues such as hallucinations and loss of normative nuance \cite{haque2025}. Wang et al. apply LLMs like GPT-4 to autonomous driving scenarios, showcasing norm-following behavior in tasks such as intersection handling and platooning \cite{wang2024can}. Ashery et al. \cite{ashery2025emergent} study the emergence of collective conventions in large-scale LLM agent groups, demonstrating how naming conventions and social biases can spontaneously arise, and how minority influence can shift prevailing norms. Rezaei et al. \cite{rezaei2025egonormia} present EGO NORMIA, a benchmark that tests LLMs’ grasp of physical-social norms in egocentric visual contexts, emphasizing the importance of situated grounding. Yuan et al. \cite{yuan2024measuring} assess LLMs’ social norm alignment across cultures, revealing both conformance and divergence across global datasets and underscoring the complexity of aligning AI with diverse societal expectations. Horiguchi et al. \cite{horiguchi2024evolution} delve further into meta-norm emergence, where agents develop higher-order rules, such as punishing those who fail to sanction norm violations, through natural language interaction, reflecting the growing sophistication of norm dynamics in AI systems.

\textbf{Focus of our work} - Our work here is aligned with the work in NorMAS community where the goal is to design norm-capable agents such as robots. A recent work by He et al. \cite{he2024norm} explores norm violation detection using simulated textual stories set in a household environment. They define a fixed set of ten norms and evaluate whether three LLMs can accurately classify norm violations based on sequences of events, benchmarking the results against human judgments. While their work focuses on binary classification of norm compliance, it does not investigate a range of normative reasoning capabilities of LLMs. In contrast, our work significantly extends this line of inquiry by introducing six variants (e.g., reasoning with scenarios representing norm adherence with and without praise, and norm violation with and without punishment) compared to just one considered by He et al. Furthermore, we assess the normative reasoning abilities of MLLMs by providing textual and image inputs, while He et al. considered text inputs only. This work particularly considers image inputs to model what the robots (agents) see in a situated environment, in addition to text-based inputs.

\section{\uppercase{Methodology}}
For evaluating the five models, three specific variables were examined, namely different \textit{scenarios}, \textit{variants} and \textit{questions} (which are described in the sub-sections below). Short text-based stories (100 words) corresponding to each combination of scenario and variant were generated using GPT-4o, which were then used to generate comic strips with four panels to visually represent the story. 

\subsection{Models}
We selected five state-of-the-art MLLMs and investigated their normative reasoning abilities. The selected models were GPT-4o, Gemini 2.0 Flash, Qwen-2.5VL (72B), Intern-VL3 (14B), and Meta LLaMa-4 Maverick. GPT-4o was accessed via OpenAI's ChatGPT platform. The other models were accessed via API calls and the results were analyzed.

\subsection{Scenarios and Variants}
In our study, we considered five different scenarios, each depicting a different norm, namely 1) knocking on a door before entering a room, 2) not littering in a park, 3) maintaining the order in a line, 4) being punctual, and 5) offering one's seat to the elderly. These scenarios depict well-known behavioral expectations for an agent in a social context representing prohibition norms (e.g., not littering) and obligation norms (e.g., knocking the door), the two common norm types investigated in NorMAS literature.

In addition, each of these five scenarios comprised six variants. In each variant, the state of the norms (adherence or violation), and the consequence for norm adherence or violation are different (as shown in Table \ref{variant_types}). These variants were designed to test the nuanced normative understanding and reasoning abilities of MLLMs. All the considered variants are described below. These variants were developed based on prior research work in NorMAS \cite{boella2006introduction,morris2019norm,savarimuthu2011norm,he2024norm,savarimuthu2024harnessing}.

\begin{enumerate}
    \item[V1:] An agent adheres to the social norm without receiving praise for doing so.
    \item[V2:] An agent adheres to the social norm and receives praise from bystanders.
    \item[V3:] An agent violates the social norm, but does not receive sanctions for violation.
    \item[V4:] An agent violates the social norm and receives one of the forms of sanctions as described below:
        \begin{enumerate}
            \item[V4a:] 
            A bystander offers gentle advice to the norm violator agent not to violate norms in the future
            \item[V4b:]
            A bystander scolds the norm violator agent
        \end{enumerate}
    \item[V5:] A sanctioning bystander agent  not only sanctions the violator of the norm but also sanctions other passive bystanders who do not sanction the violation of the norms. This form of sanction is called meta-punishment and these norms are called metanorms (norms governing violation of norms). 
\end{enumerate}

\newcommand{\cmark}{\checkmark} 
\newcommand{\dash}{--}

The six variants presented above represent three categories of norm adherence or violation: 
\begin{enumerate}
\item[C1:] Norm adherence (V1 and V2)
\item[C2:] Norm violation (V3, V4a, V4b)
\item[C3:] Metanorm (V5)
\end{enumerate}

Table \ref{variant_types} shows whether a variant of the story includes specific dimension of normative reasoning (\textcolor{blue}{'\cmark'} = included in that variant, and '\dash'  = not applicable for that variant).

\begin{table*}[!h]
\centering
\begin{tabular}{|p{1cm}|p{2cm}|p{2cm}|p{2cm}|p{1.9cm}|p{2cm}|p{1.9cm}|}
\hline
Variants & Norm Adherance & Norm Adherance Praise & Norm Violation & Gentle Advice Against Violation & Scolding to Correct Norm Violation & Meta Punishment\\
\hline
1 & \textcolor{blue}{\cmark} & \dash & \dash & \dash & \dash & \dash \\
\hline
2 & \textcolor{blue}{\cmark} & \textcolor{blue}{\cmark} & \dash & \dash & \dash & \dash \\
\hline
3 & \dash & \dash & \textcolor{blue}{\cmark} & \dash & \dash & \dash \\
\hline
4a & \dash & \dash & \textcolor{blue}{\cmark} & \textcolor{blue}{\cmark} & \dash & \dash \\
\hline
4b & \dash & \dash & \textcolor{blue}{\cmark} & \dash & \textcolor{blue}{\cmark} & \dash \\
\hline
5 & \dash & \dash & \textcolor{blue}{\cmark} & \dash & \textcolor{blue}{\cmark} & \textcolor{blue}{\cmark} \\
\hline
\end{tabular}
\caption{Norm adherence and violation aspects considered in each variant}
\label{variant_types}
\end{table*}

\subsection{Questions}
For evaluating the MLLM response for a given story, we used a set of eight questions to analyze how accurately it can identify norm violations from texts and images. The questions are presented in Table \ref{tab:my_label}.

\begin{table*}[t]
\centering
\begin{tabular}{|c|c|c|}
        \hline
         Question No. & Question & Expected Answer \\
        \hline
        1 & What is the norm? & Brief descriptive answer \\
        \hline
        2 & Who is the subject of the norm? & Brief descriptive answer \\
        \hline
        3 & Is the norm adhered to at the beginning of the scenario? & Yes/No
        \\
        \hline
        4 & Is the norm adherence praised in the scenario? & Yes/No
        \\
        \hline
        5 & Is there norm violation at the beginning of the scenario? & Yes/No
        \\
        \hline
        6 & Is the norm violator educated through gentle and calm advice? & Yes/No
        \\
        \hline
        7 & Is the norm violator being scolded for norm violation? & Yes/No
        \\
        \hline
        8 & Does a punisher punish those who do not punish norm violators? & Yes/No
        \\
        \hline
    \end{tabular}
    \vspace{10pt}
    \caption{Questions}
    \label{tab:my_label}
\end{table*}

In Table \ref{tab:my_label}, questions 1 and 2 receive short descriptive answers from MLLMs. For questions 3-8, MLLMs provide Yes or No as answers.  Evaluating answers to these questions will reveal how well MLLMs understand and interpret social norms from textual or image-based stories. By covering a range of dimensions, from identifying the norm and its subject in the first two questions, to recognizing adherence or violation, and further probing into social consequences, the questions helped assess whether MLLMs can go beyond surface-level pattern recognition, to engage in deeper social reasoning. Our questions included the scrutiny of understanding of normative reasoning of MLLMs based on implicit cues, moral accountability, and the dynamics of praise, sanctions, and meta-enforcement that govern human behavior. Such a multifaceted approach is essential for assessing the robustness and reliability of MLLMs in socially grounded real-world applications where interpreting human norms accurately is critical, particularly in scenarios involving human-robot interactions.

\subsection{Text-Based Analysis}
\begin{figure*}[!h]
\centering
\begin{minipage}{.5\textwidth}
  \centering
  \includegraphics[width=\linewidth]{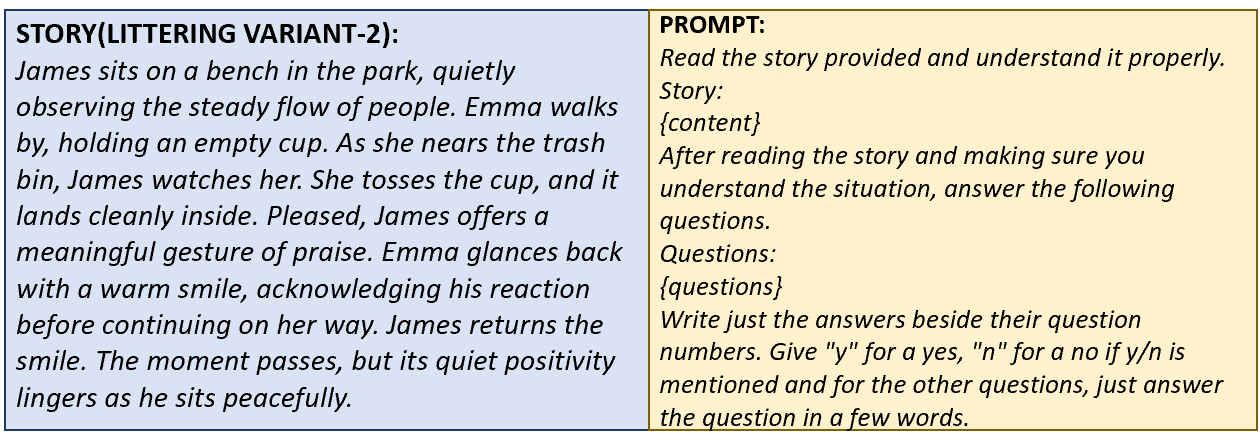}
  \captionsetup{skip=6pt}
  \captionof{figure}{Response Generation For Text}
  \label{text_response}
\end{minipage}%
\begin{minipage}{.5\textwidth}
  \centering
  \includegraphics[width=\linewidth]{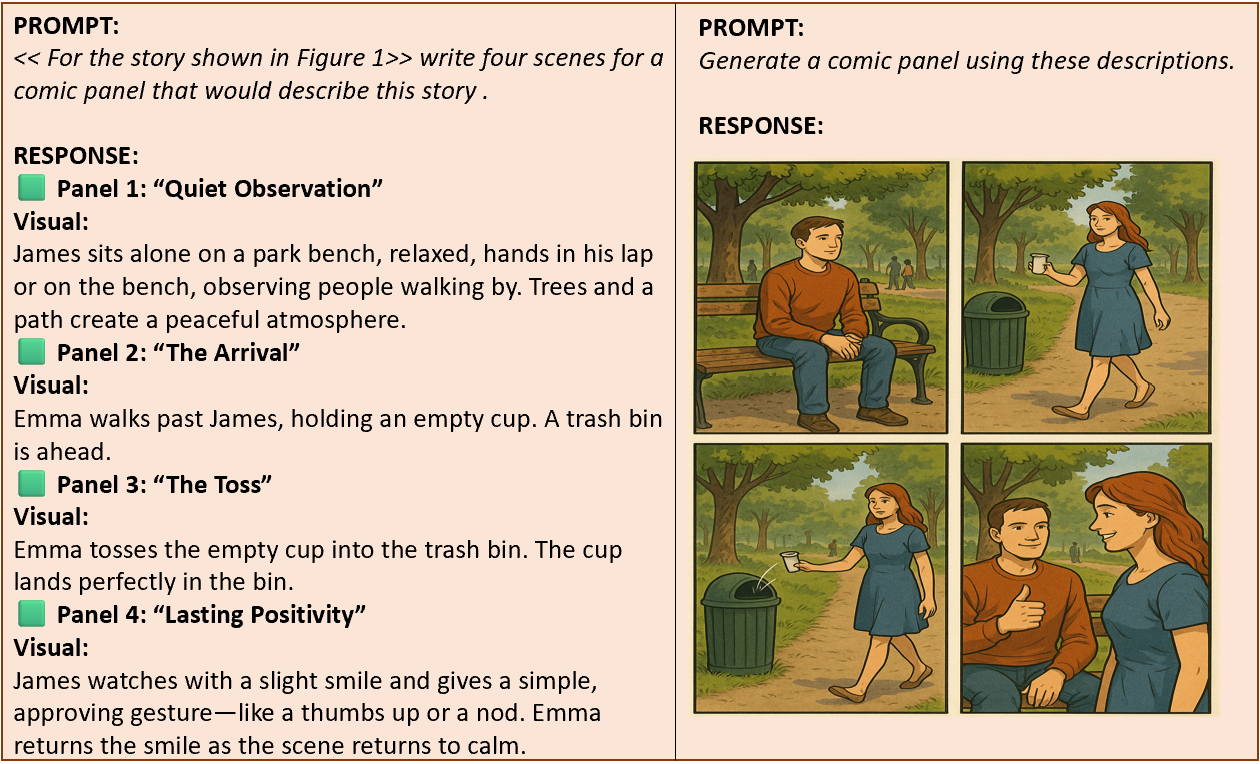}
  \captionsetup{skip=6pt}
  \captionof{figure}{Image Generation}
  \label{image_generation}
\end{minipage}
\end{figure*}

As indicated in Section 3.1, thirty text-based stories were analysed. An example  text-based story is presented on the left of Fig. \ref{text_response}, which corresponds to variant 2 (i.e. where there is adherence to a social norm and the adherer receives praise) for not littering. 

To establish the ground truth, two human evaluators (the first and the second author of the paper) read all the stories and answered the eight questions to establish the ground truth based on consensus. To address the robustness of our evaluation, we also included two external evaluators. One evaluator evaluated all the six variants of the textual stories for the Knocking scenario and answered the same set of questions which had a 100\% agreement with the authors' ground truth. The same participant was asked to answer the same questions based on analysing the images of the Knocking scenario which had a 95.83\% match with the authors' ground truth. Additionally, a second external evaluator evaluated the six variants of the Maintain-Line scenario and answered the same set of questions which had a 97.91\% and 95.83\% agreement with the authors' ground truth, for text and image based scenarios respectively. The Cohen's Kappa score for Knocking scenario was 0.93 and for Maintain-Line scenario was 0.90. This shows almost perfect agreement \cite{mchugh2012} between evaluators.

To evaluate the performance of the five MLLMs, they were presented with all story variants, and they were prompted to answer the questions shown in Table \ref{tab:my_label}, using the prompt shown on the right side of Fig. \ref{text_response}. Comparisons were made for each combination of scenario, variant and question, against the established human ground truth, and every answer was marked as either correct or incorrect.


\subsection{Image-Based Analysis}
Comic strip like images were generated for the same 30 text-based stories. To create these comic-strip like images for the story, GPT-4o was firstly prompted to create a written description of four scenes for each story. This description was provided as the input to GPT-4o to generate a depiction of the story in four panels of the comic strip. An example of the generation process of the images is displayed in Fig. \ref{image_generation}, with the left part showing the description of four scenes in words and the right part showing the comic strip generated by GPT-4o comprising four scenes. A handful of these images had to be edited for coherence. Textual stories and generated images can be found in this GitHub repository\footnote{\href{https://tinyurl.com/jfp6xzjn}{https://tinyurl.com/jfp6xzjn}}.


Once these images were generated, the models were prompted one by one to generate the story that they understood from that particular image and then answer the eight questions based on their understanding of that generated story. The response from the models were compared against ground truth, and the MLLM responses were marked as either correct or incorrect.

\section{Results}
We analysed the responses from the MLLMs which were marked as correct or incorrect. The accuracy score was computed using the formula: correct predictions / correct predictions + incorrect predictions and these results were visualised in the form of graphs. Fig. \ref{fig:text-boxplots} and Fig. \ref{fig:image-boxplots} each present a comprehensive set of 12 box-plots arranged in a 4$\times$3 grid. These visualizations are designed to compare the performance of different MLLMs against each other, based on three variables:  questions, variants, and scenarios. The y-axis score shows the accuracy of the models. 

\subsection{Text-Based Analysis}
\begin{figure*}[!h]
    \centering
    \includegraphics[width = 0.85\linewidth]{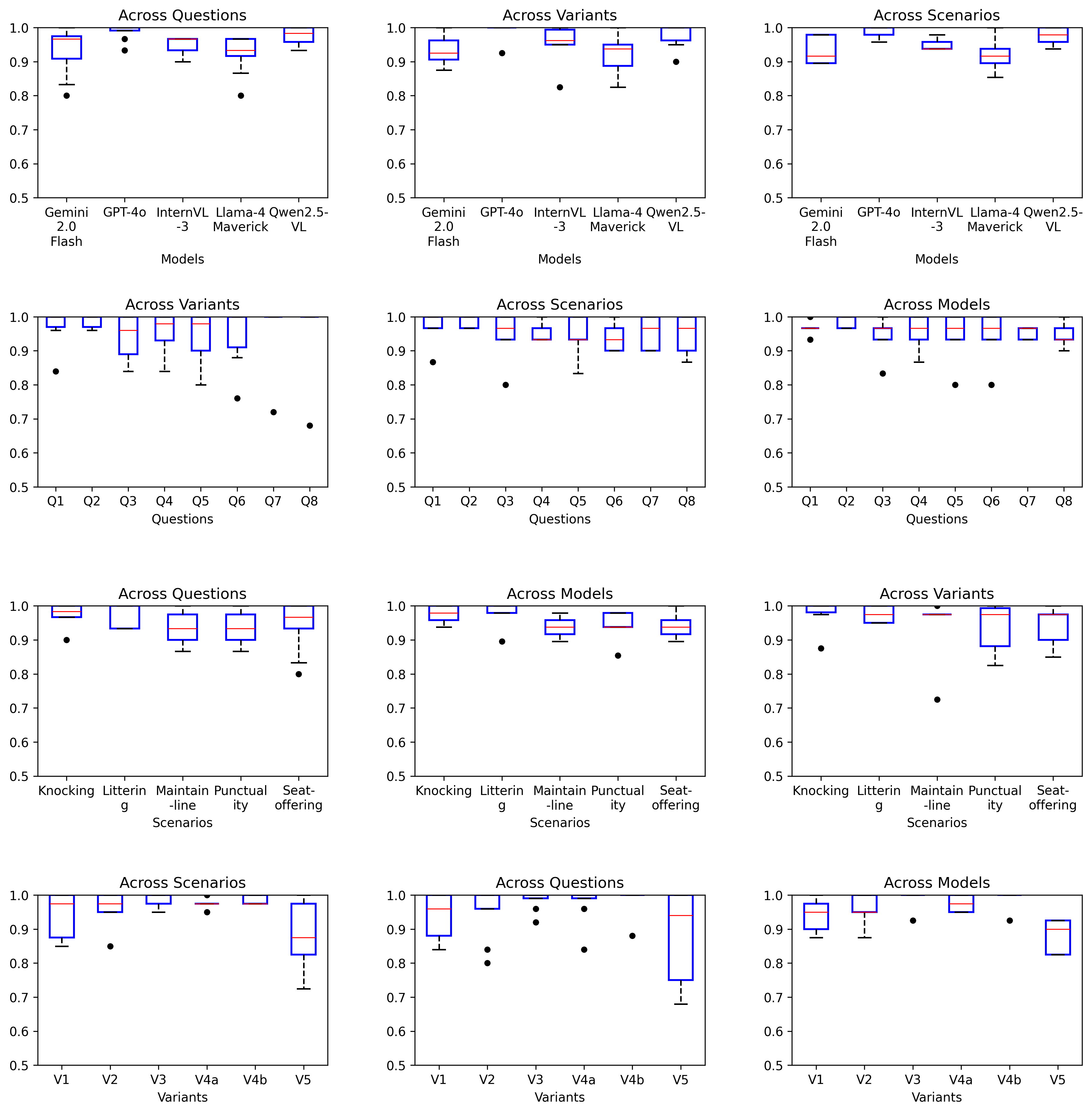}
    \caption{Text Accuracies: Boxplot}
    \label{fig:text-boxplots}
\end{figure*}

Fig. \ref{fig:text-boxplots} presents the results for text-based stories. The top row focuses on the distribution of accuracies observed from the models' outputs across questions, variants and the scenarios. In the first plot (at position 1,1 - i.e., row 1, column 1), shows the models' performance aggregated over all questions. GPT-4o and Qwen2.5-VL demonstrate strong performance ($>$96\% accuracy) with tight interquartile ranges, indicating stable behavior across the question set. In contrast,  LLaMa-4 Maverick exhibits slightly lower median score and greater variability compared to other models, suggesting inconsistencies in its comprehension or reasoning capabilities. The second plot (at position 1,2) aggregates model performance across all variants. Once again, GPT-4o and Qwen2.5-VL had the top median scores. The third plot (at position 1,3) illustrates model performance across different scenarios. Again, GPT-4o and Qwen2.5-VL continue to display strong and stable scores, while LLaMa-4 Maverick and Gemini 2.0 Flash show lower medians and slightly larger variations.

The second row presents the accuracy with which the questions were answered. The first plot (at position 2,1) evaluates how accurately each question was answered when assessed across all variants. Questions Q1 and Q2 show the least variability and good median scores suggesting the best performance (\textgreater96\%), with Question Q2 not having any outliers. Questions Q3-Q6 showed similar performance in terms of their distributions. On the other hand, questions Q7 and Q8 had good median scores but showed very low outlier scores, indicating that the models could have had difficulty in answering these for some variants. The second plot (at position 2,2) evaluates the accuracy of answering questions across the scenarios, which reveals the performance for all questions was on par with each other, of which, question Q2 has the least variability without outliers, suggesting that answering the question was consistent across scenarios. The third plot (at position 2,3) evaluates the accuracy of answering questions across the models. It reveals that questions Q1 and Q2 were answered with highest accuracy (\textgreater96\%), revealing they were easier for all the models to answer, compared to the rest.

The third row dives into scenario-level comparisons. The first plot (at position 3,1) aggregates scenario performance across all questions, revealing 'Knocking' and 'Littering' perform comparatively better than the other three scenarios. Some outliers are present, pointing to some questions being generally difficult to understand or answer. The second plot (at position 3,2) evaluates scenarios across all models, showing a generally high level of performance, with a few outliers in the case of Littering and Punctuality. The third plot (at position 3,3) assesses how scenarios perform across different variants, pointing to increased variability in the scenarios involving `Punctuality' and `Seat-offering' norms, suggesting some variants may be more difficult to answer for compared to others. Additionally, there are a few low score outliers (e.g., in the `Maintain-line' scenario).

The bottom row focuses on the accuracy of understanding the different variants. Variant V5 consistently underperforms across scenarios, questions, and models, with lower medians and higher variability, as evident from the first plot (at position 4,1). This suggests potential structural complexity in variant V5 (i.e., it involves multiple levels of reasoning). However, questions for variant V2 seem to be relatively difficult to answer as shown by the outliers in the second plot (at position 4,2) in the bottom row. The third plot (at position 4,3) reinforces the poor performance of the models for variant V5, indicating a difficulty in understanding the complexity of the variant.

\subsection{Image-Based Analysis}
\begin{figure*}[!h]
    \centering
    \includegraphics[width=0.85\linewidth]{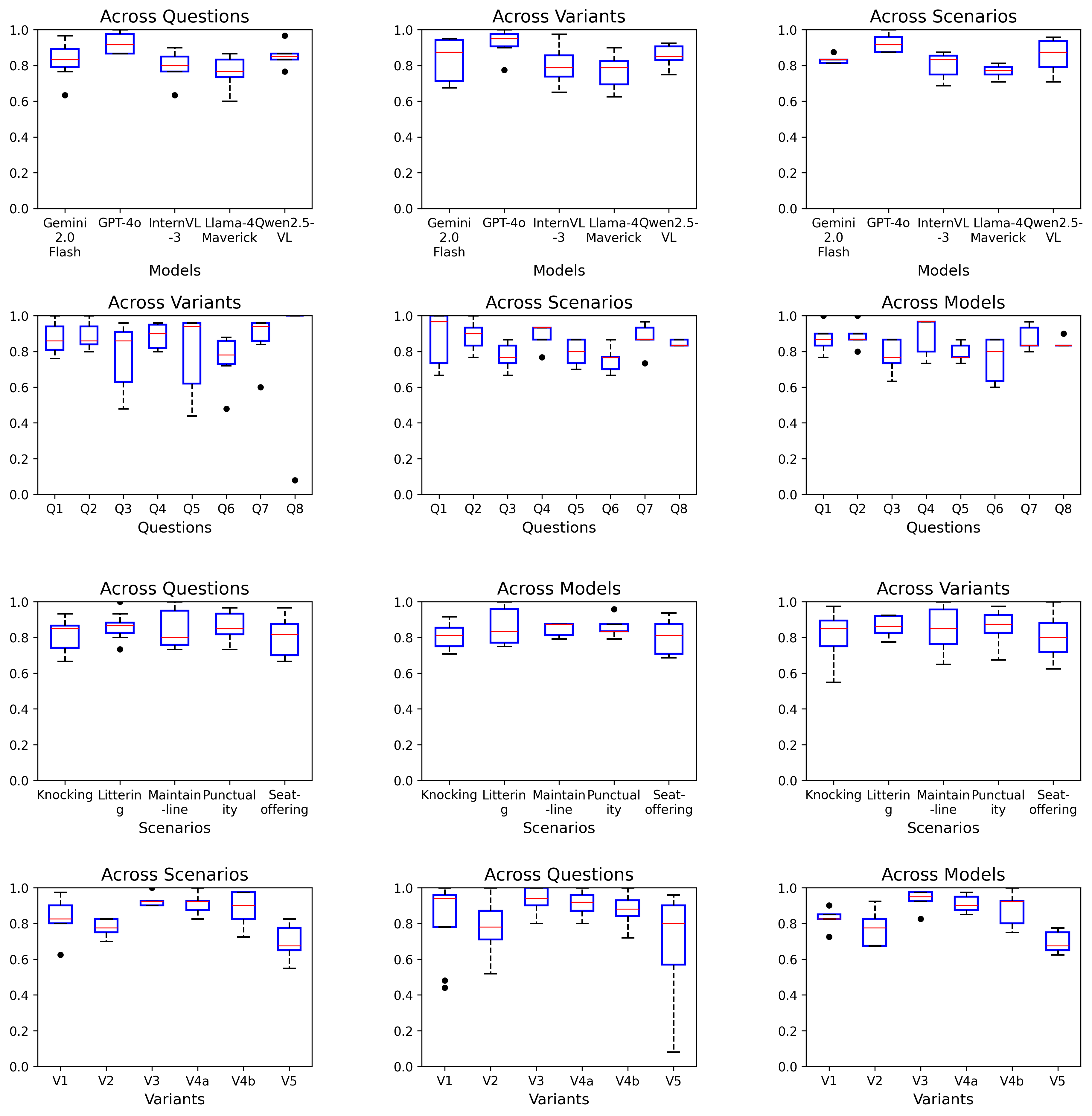}
    \caption{Image Accuracies: Boxplot}
    \label{fig:image-boxplots}
\end{figure*}

Fig. \ref{fig:image-boxplots} presents 12 boxplots that depict the performance of the models across scenarios, variants and questions when image inputs are presented. The first row reports the distribution of accuracies achieved by the models, with the first plot (at position 1,1) depicting how accurately the models answered questions. It can be seen that GPT-4o performed the best with a higher median score compared to the other models, closely followed by Qwen2.5-VL and the others. Qwen2.5-VL, however, has the lowest variability, indicating a similar performance across all questions except a few outliers. The second plot (at position 1,2) depicts the distribution of how each model performed for all the variants and GPT-4o outperforms all other models. Furthermore, Gemini 2.0 Flash shows the largest interquartile range, indicating a more varied performance when compared to other models. The last plot (at position 1,3) shows the performance of the MLLMs across the different scenarios, where we see a similar dominance from GPT-4o. Gemini 2.0 Flash and LLama-4 Maverick show tight interquartile ranges compared to others, however they seem to also exhibit the lowest median scores out of all models in the plot.

The second row reveals the performance in answering the individual questions. The first plot (at position 2,1) shows the performances of all the questions across all the six variants of images. Questions Q3 and Q5 have larger interquartile ranges compared to other questions, indicating high variability in how accurately the questions were answered for different variants. The second plot (at position 2,2) reveals how accurately the questions could be answered across different scenarios, which indicates questions Q2, Q4 and Q7 show similar results, which are much higher compared to the others. Furthermore, question Q1 shows the largest spread, demonstrating a high variability in answering this question, indicating an increased difficulty in understanding the different scenarios. The last plot (at position 2,3) depicts how accurately the questions could be answered by the models. Questions Q1, Q2 and Q4 are answered with high median scores, while Q3 and Q6 show larger interquartile ranges with low median scores, indicating that models struggled more with these questions.

In the third row, scenario-level accuracy is analyzed by questions, models, and variants. The first plot (at position 3,1) shows that the accuracy of answering questions pertaining to scenarios like ``Littering” and ``Punctuality” had higher median scores and narrower spreads, indicating better understanding of these normative scenarios. In contrast, ``Knocking”, ``Seat-offering” and ``Maintain-line” had wider spreads and lower medians, suggesting greater complexity. The second plot (at position 3,2) highlights similar (lower) performance by models. Higher spreads in "Littering" and "Seat-offering" point towards the models having less understanding of the scenario. The third plot (at position 3,3) shows similar accuracy levels for variants ``Punctuality” and ``Littering”. The others had higher variability, indicating that some variants might have been more challenging to comprehend.

The final row analyzes accuracy across image variants. In the first plot (at position 4,1), variants V3 and V4a perform best, with high medians and narrow spreads, suggesting the images for these scenarios clearly conveyed the intended norms. In contrast, V2 and V5 show lower median and wider spread, indicating difficulty in interpretation. The second plot (at position 4,2) reinforces this: V3, V4a, and V4b perform consistently across questions, while V2 and V5 show greater variability and outliers. The third plot (at position 4,3) highlights that the models generally understand variants V3 and V4a better compared to  other variants. Meanwhile, variants V2 and V5 show worse performances with low median scores indicating high complexity of these variants resulting in relatively low level of understandability.

\subsection{Statistical comparisons of Text and Image Analysis Results}
\begin{figure*}[!h]
    \centering
    \includegraphics[scale=0.45]{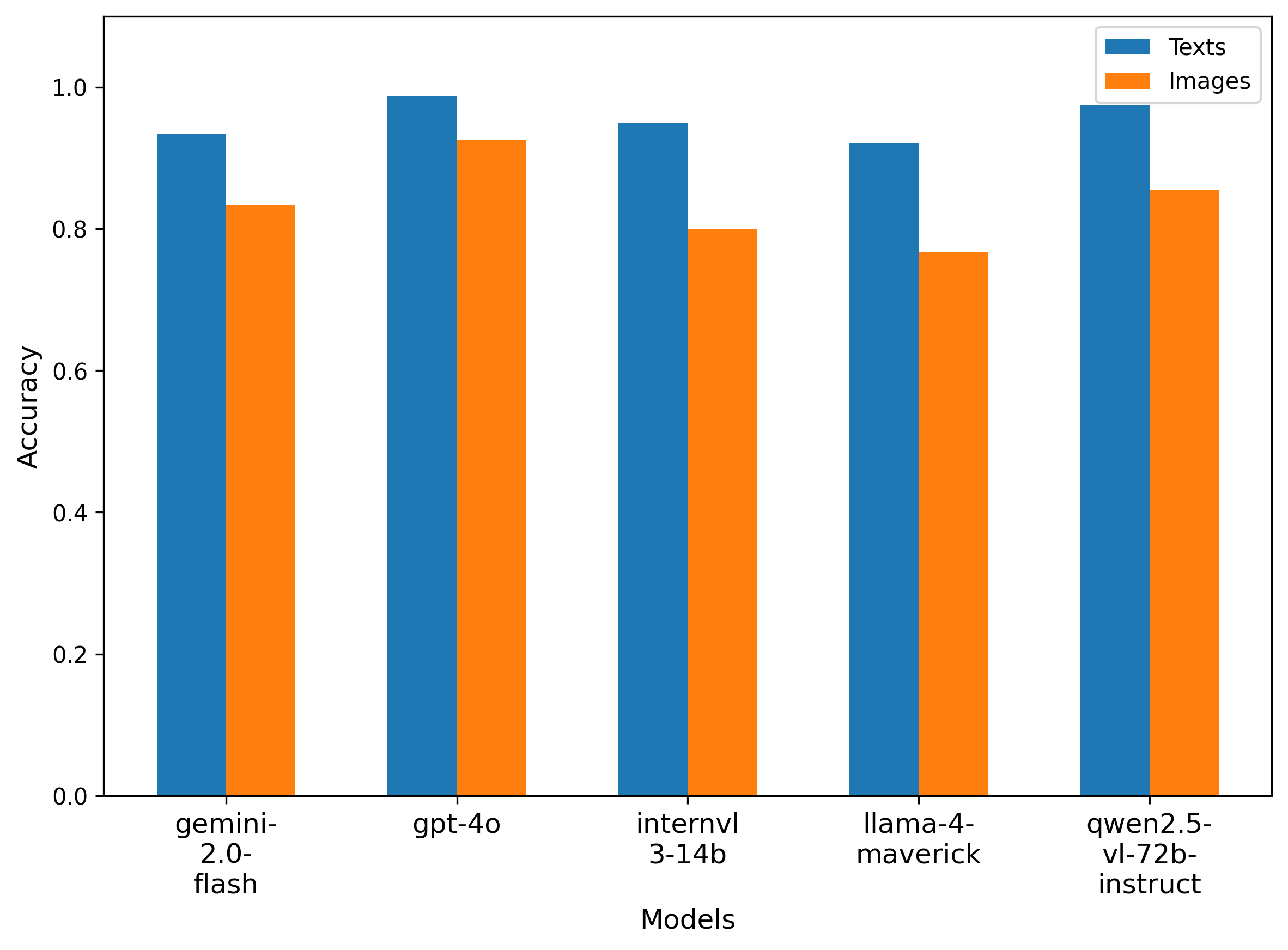}
    \caption{Model Accuracies-Text vs Images}
    \label{fig:bar-graph}
\end{figure*}

Fig. \ref{fig:bar-graph} compares average model accuracy scores of models and it can be observed that MLLMs consistently performed better when analyzing norms in textual stories than images. A paired-samples t-test indicated that the five models performed significantly better on text (M = 0.95, SD = 0.09) than on images (M = 0.84, SD = 0.17), t(149) = 8.91, p $<$ 0.001, with a large effect size (Cohen’s $d_z$= 0.82).

\begin{figure*}[!h]
    \centering
    \includegraphics[scale=0.25]{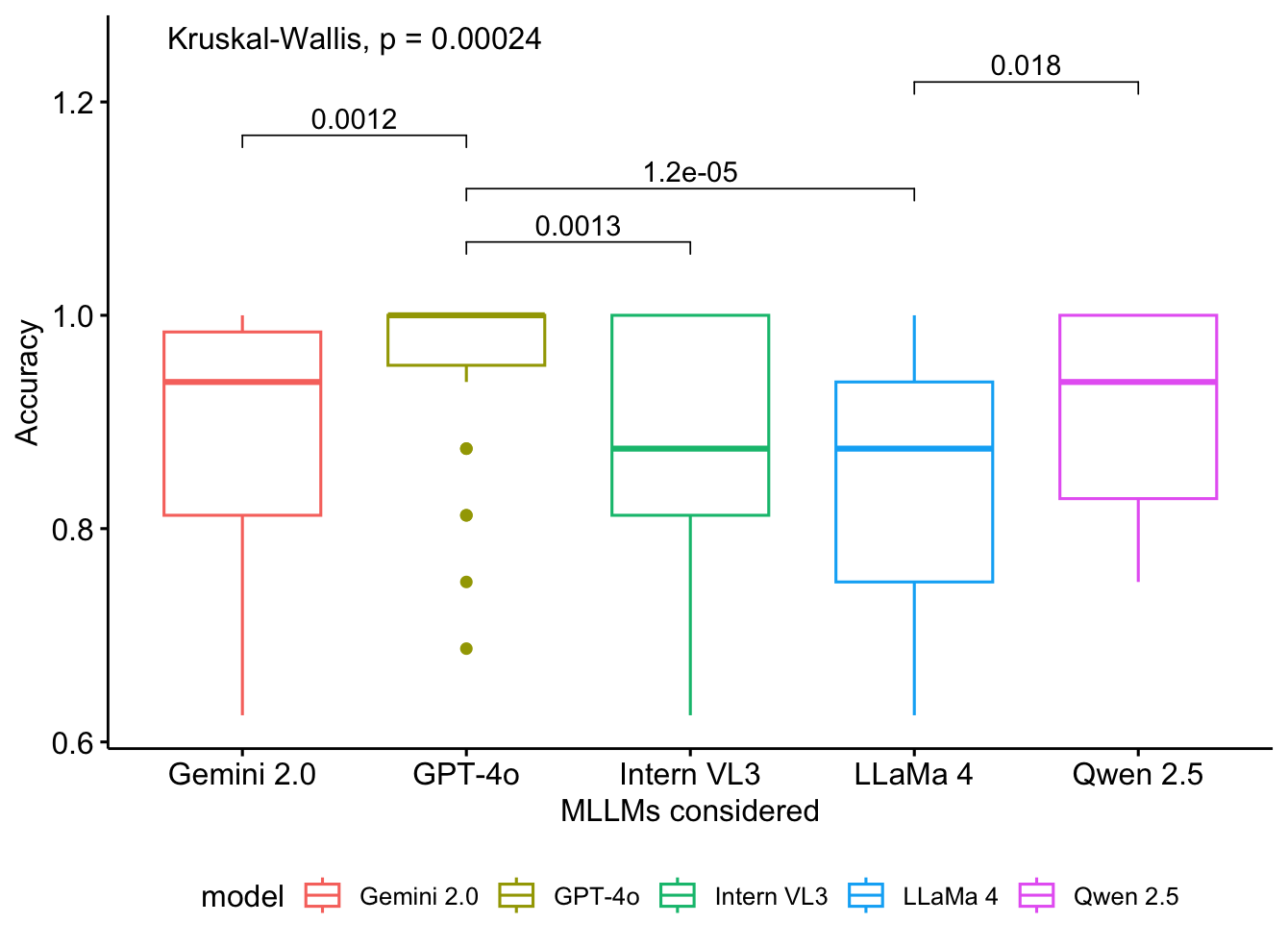}
    \caption{Box plot of model accuracies across five models in analysing text and images}
    \label{fig:significance_results}
\end{figure*}


When comparing combined results of text and image analysis across all five algorithms, the Friedman test revealed a significant difference among the algorithms, $\chi^2$(4) = 36.32, p $<$ 0.001. Follow-up post-hoc pairwise Nemenyi tests and Wilcoxon signed-rank tests showed that GPT-4o significantly outperformed LLaMA 4 (p $<$ 0.001, r = 0.81 (with r $>$ 0.5 signifying a large effect size)), Intern VL (p $<$ 0.01, r = 0.68), and Gemini 2.0 Flash (p $<$ 0.01, r = 0.67). Additionally, Qwen 2.5 performed significantly better than LLaMA 4 (p $<$ 0.05, r = 0.55). All other pairwise comparisons were not statistically significant. The exact p-values for the four significant results are shown above the boxplots in Figure \ref{fig:significance_results}. These results indicate that GPT-4o achieved the highest performance, with Qwen2.5-VL ranking second.

\begin{figure*}[!h]
   \centering
    \includegraphics[scale=0.35]{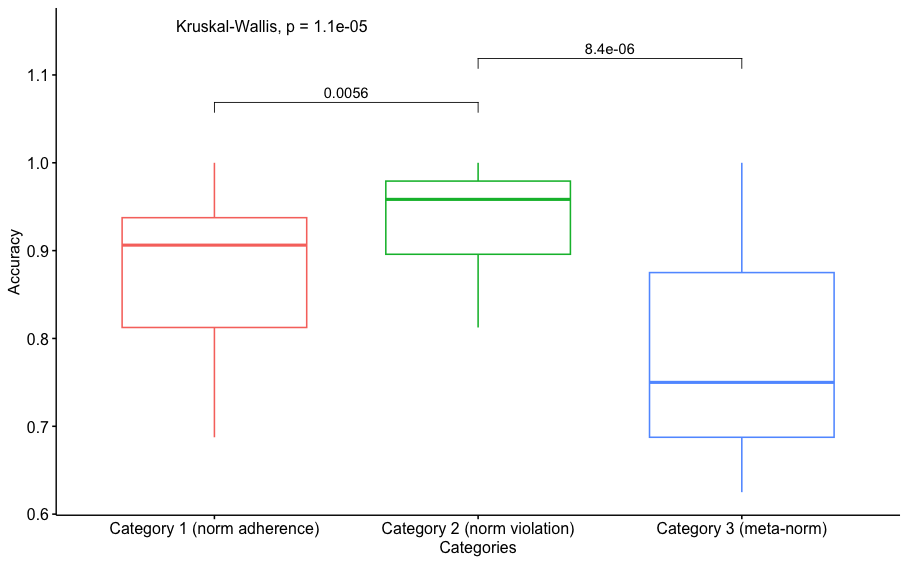}
    \caption{Box plot of model accuracies across three categories}
    \label{fig:category_comparison}
\end{figure*}


When comparing combined results of text and image analysis across three categories of variants as indicated at the end of Section 3.1 (category 1 - focussing on adherence, category 2 - focussing on norm violation and category 3 - focussing on meta norms), Friedman test revealed a significant difference among the three categories, $\chi^2$(2) = 21.38, p $<$ 0.001. Follow-up post-hoc pairwise Nemenyi tests and Wilcoxon signed-rank tests showed statistically significant differences between two pairs of categories. MLLMs performed better in detecting norm violations (category 2) than norm adherences (category 1), with p $<$ 0.01, r = 0.61. Also, MLLMs performed better in detecting norm violations (category 2) than detecting norms  metanorms (category 3), with p $<$ 0.01, r = 0.8. This result is visualised in Figure \ref{fig:category_comparison} which shows the p-values for siginicant results at the top of the image. This result also shows that reasoning about metanorms (category 3) was the most challenging for the models, with a median accuracy of 75\% (which is lower than the median scores for category 3 and category 2 which are 90.6\% and 95.8\% respectively). 


\section{\uppercase{Discussion}}
Unlike prior studies, our work presents a comprehensive framework for evaluating normative reasoning by incorporating rich, in-depth social scenarios considering five norms (scenarios), with several variants each (six for both texts and images) and systematically designed question set comprising eight questions to rigorously evaluate various facets of norm adherence and violation. Furthermore, we extend the scope of analysis to include both texts and images, enabling evaluation of multimodal LLMs, an area that has remained largely unexplored in earlier research. 

To strengthen the validity of our experimental setup we conducted an additional experiment. This was to mitigate the potential bias arising from the use of GPT-4o in reasoning from its own generated images. For this experiment we generated image-based stories for the Littering scenario variants (six of them) using the model Seedream 4.0\footnote{https://seedreamai.io/}, and analysed the results from all the five models. Our results revealed that in this test, the models performed similar to other tests, with GPT-4o outperforming all other models hence, ruling out bias.

Our empirical findings reveal several key observations. First, all evaluated LLMs exhibit stronger performance in text-based normative reasoning compared to analysing images. This suggests that existing LLMs possess more advanced capabilities in textual inference and reasoning than in visual understanding of social contexts. Among the models, GPT-4o consistently outperforms others, achieving 98.75\% accuracy on text-based inputs and 92.5\% accuracy on image-based inputs. These results highlight GPT-4o's superior generalization across modalities while also underscoring the broader trend of text-dominant reasoning strengths among current multimodal LLMs. On the other hand, Meta LLaMa-4 Maverick performs the worst, achieving 92\% accuracy on text-based inputs and 76.66\% on image-based inputs. When considering free-to-use LLMs, Qwen2.5-VL performed the best achieving 97.5\% accuracy on text-based inputs and 85.41\% on image-based inputs.

\begin{figure*}[!h]
    \centering
    \includegraphics[scale=0.09]{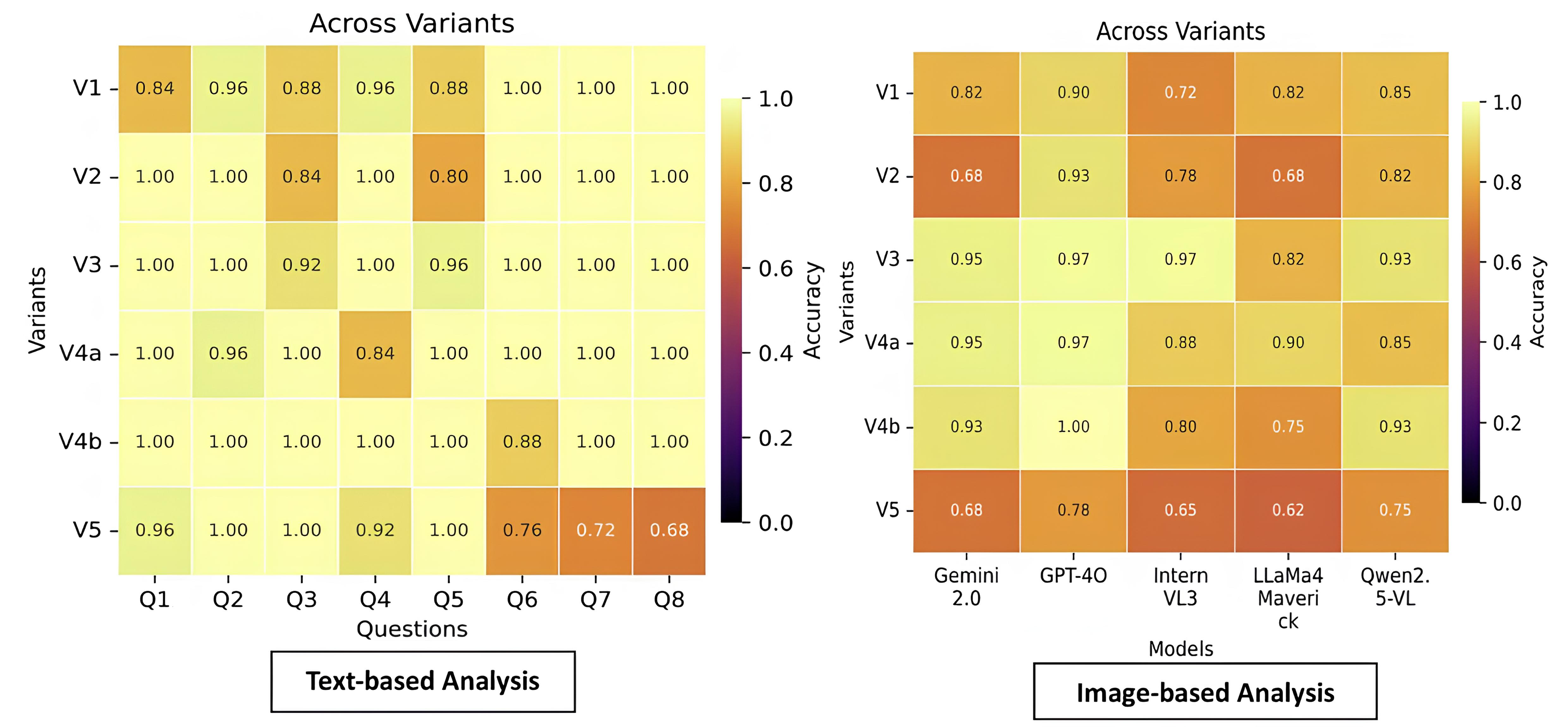}
    \caption{Heatmaps of results from variants across questions for text and image analysis}
    \label{fig:heatmaps}
\end{figure*}

Now we turn to specific instances where the models did not perform well on text-based stories. First, we observed that LLMs performed the worst in the case of variant V5. The difficulty in understanding this variant could have occurred due to the reasoning about metanorm including multiple deeper levels of analysis (level 1 - to identify whether someone had violated the norm, level 2 - to understand whether someone is punishing the norm violator and level 3 - to identify other bystanders (if any) who did not sanction the norm violator). Second, in the case of norms considered (i.e., scenarios), the LLMs showed the best performance for the Littering and Knocking norms. A heatmap based analysis for the text analysis, showing performance trends is included in the GitHub repository\footnote{\href{https://tinyurl.com/45yy62bn}{https://tinyurl.com/45yy62bn}}. The image on the left of Figure \ref{fig:heatmaps} is an example of one of the heatmaps for text-based analysis. It shows that variant V5 has the worst performance among all variants, and the performance is especially worse for Q8. This reveals that a lot of models fail to correctly evaluate Q8 in the case of variant V5 (see last row), which is the question targeting that specific variant. We also observed a similar trend for image-based stories (not shown here).

Turning to specific instances where models fared poorly in answering norm-related questions based on images, first, we find that questions 3, 5 and 6 were somewhat challenging. We believe this is due to a) difficulty in understanding the progress of the scenes in a comic strip or b) misunderstanding the questions. Secondly, the LLMs showed a consistent performance across all scenarios, but "Seat-offering" appears to be the hardest to understand based on low medians and wide interquartile ranges. This could be because seat offering may not have been well depicted in the comic strips. Third, variant 5 was the worst performing variant followed by variant 2. This performance for variant 5 could be due to the inherent complexity and the layers of reasoning required (as discussed earlier). For variant 2, which contains praise for norm adherance, the MLLMs might have found it difficult to understand the action of praise from the images. Additionally in a few cases (e.g., variant 4b) there was a plausible alternate story that could be generated from the images which points to a different norm (e.g., politeness norm during interactions). A heatmap based analysis for images, showing performance trends is included in the GitHub repository\footnote{\href{https://tinyurl.com/45yy62bn}{https://tinyurl.com/45yy62bn}}. The right of Figure \ref{fig:heatmaps} contains an example of one of the heatmaps for image-based analysis. It shows that variant V5 (see last row) performed the worst among all variants. It can also be seen that the model LLaMa-4 Maverick performed the worst among all other models (see column 4).

Having said the above, the overall average accuracy across models for text-based and image-based norm reasoning was 95.33\% and 83.58\% respectively highlights the promise of the use of MLLMs for norm reasoning. Several promising directions remain for expanding this work further. First, having explored both text-based and image-based reasoning, a natural progression involves extending the framework to video-based analysis \cite {fu2025video}, comprising multimodal inputs (audio, text and images). Second, in this study, we used zero-shot prompting to assess innate model capabilities without in-context examples or fine-tuning. Future work could explore fine-tuning and Retrieval-Augmented Generation (RAG) for domain specialization. Also, advanced strategies like Tree-of-Thought can be tried to improve reasoning on complex moral and social dilemmas. Third, the normative categories studied here (obligations and prohibitions), could be broadened to include permissions, institutional rules, and culture-specific norms, supporting better generalization across diverse sociocultural contexts. Fourth, the use of multiple agents can be considered to analyse multi-modal aspects (one specializing in text, other in audio, and yet another in images) and also these agents can query an an ensemble of different algorithms (e.g., using majority voting based on results from different MLLMs for image analysis) to answer specific questions. However, trim, space and cost tradeoffs should be considered. Fifth, an important direction is the real-world evaluation of norm-aware MLLMs in embodied agents like social robots or virtual or augmented reality-based assistants. In such settings, the ability to interpret and act on norms accurately becomes critical, especially when norms are implicit or culturally contingent. For instance, GPT-4o’s multimodal reasoning could allow robots to combine speech and gesture interpretation to recognize and adopt social norms in everyday interactions, directly enhancing safety, trust, and user acceptance. Lastly, future work could explore how models learn norms dynamically through interaction, via human feedback, demonstrations, or reinforcement learning, enabling the development of adaptive agents that internalize evolving social expectations over time.

\section{\uppercase{Conclusion}}
This work investigated the competency of MLLMs in normative reasoning from both text-based and image-based inputs based on five norms with different norm adherence and violation approaches (variants). Of the five MLLMs considered (GPT-4o, Gemini 2.0 Flash, Qwen-2.5VL, Intern-VL3 , and LLaMa-4 Maverick), GPT-4o showed the most promising results with 98.75\% and 92.5\% accuracy on text and image-based inputs. Statistically, the model performs better than three other models (Gemini 2.0 Flash, Intern-VL3 and LLaMa-4 Maverick). MLLMs showed better performance in analyzing norms from text than images. Additionally, our results demonstrated that MLLM models are better at reasoning with simple norms than complex norms (i.e., metanorms). These results demonstrate that, while MLLMs show promise in reasoning about norms through textual or visual inputs, there is scope for future improvements (e.g., reasoning about metanorms). Qwen2.5 VL, a free to use MLLM, can be employed by researchers and practitioners, as an alternative to GPT-4o to create socially aware robots, that can identify norm violations and punish norm violations, thus facilitating social order within these societies.

\bibliographystyle{apalike}
{\small
\bibliography{example}}



\end{document}